\documentclass{camera}
\usepackage{graphicx}  

\begin{document}

%
\title{Gamma-ray burst groups observed by BATSE, Beppo-Sax and {\it Swift}}

%
\author{I. Horv\'ath$^{1}$, L.G. Bal\'azs$^{2}$, Z. Bagoly$^{3}$, 
P. Veres$^{1,3}$ \and D. Sz\'ecsi$^{3}$}

%
\organization{$^1$ Bolyai Military University, Budapest 
$^2$Konkoly Observatory, Budapest 
$^3$E\"otv\"os University, Budapest }

\maketitle

\begin{abstract}
 Short and long bursts were identified by the BATSE team in the early 90's. 
A decade ago there were some suggestions about the 
intermediate duration type of bursts. 
We are going to summarize recent analysises of the 
duration distributions of the
Beppo-Sax and {\it Swift} data.
Our conclusion is all  the three satellites (CGRO,
{\it Swift}, Beppo-Sax) can see the third type of the GRBs.
The properties of the group members are very similar
in the different data sets.

\end{abstract}

%
 \section{Introduction}
Series of papers studied the classification of the Gamma-ray bursts' 
(GRBs). They mainly agreed among these
misterious phenomena not just short (hard) and long bursts exist
but also a third type of GRBs.
In this paper we are going to summarize these analyses
mainly done by the authors.

The discovery of the third type of GRBs goes back as early
as 1998 \cite{muk98,hor98}. After that many research groups studied
the BATSE bursts' sample and concluded the third
group of the GRBs statistically exists
\cite{hak00,bala01,rm02,hor02,hak03,bor04,hor06,chat07}.
Later several studies were published analysing
different data sets.
We are going to summarize some results
according to the CGRO, {\it Swift} and Beppo-Sax
satellites.


 \section{The BATSE sample}

The Compton Gamma-Ray Observatory (CGRO) finished its
mission in 2000 and the final BATSE catalog was published.
The duration distribution of this sample can be well
fitted with three Gaussian. The significance
that the third component is needed is 99.5\%
\cite{hor02}. 
Several papers confirmed similar or even better
significance \cite{bala01,rm02,hak03,hor06,chat07}.

\begin{figure}
\includegraphics[angle=0,width=12cm]{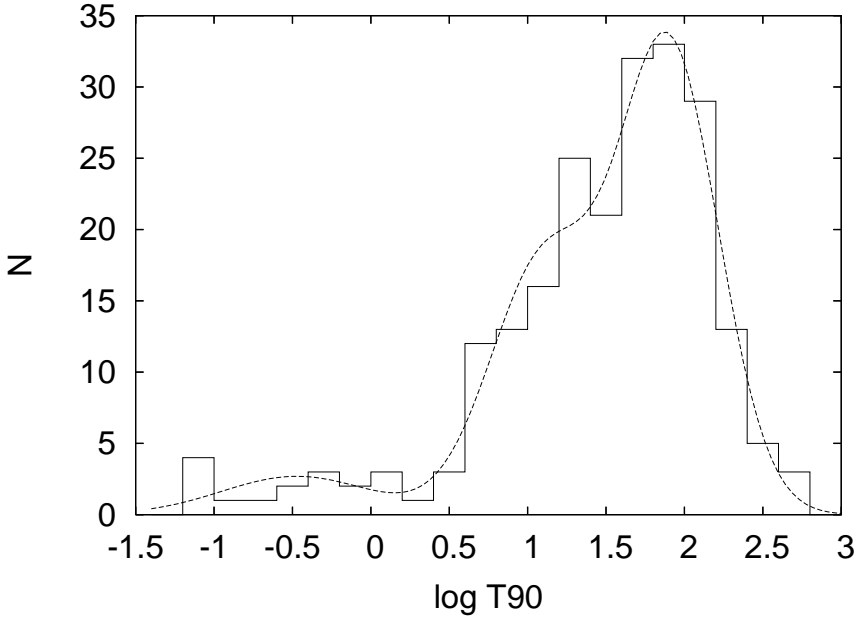}
\caption{The duration distribution of the {\it Swift} bursts and a three Gaussian fit.}
\label{fig02} 
\end{figure}

\section{The {\it Swift} sample}

The First {\it Swift}-BAT Catalog \cite{sak08} was published in 2008.
In the catalog there are 237 GRBs, of
which 222 have duration information. 
The duration distribution of this sample can be seen
in Figure 1. Also a three Gaussian fit
is plotted in the figure. The significance of
the third component is  99.4 \%
(see \cite{hor08}).

\begin{figure}
\includegraphics[angle=0,width=12cm]{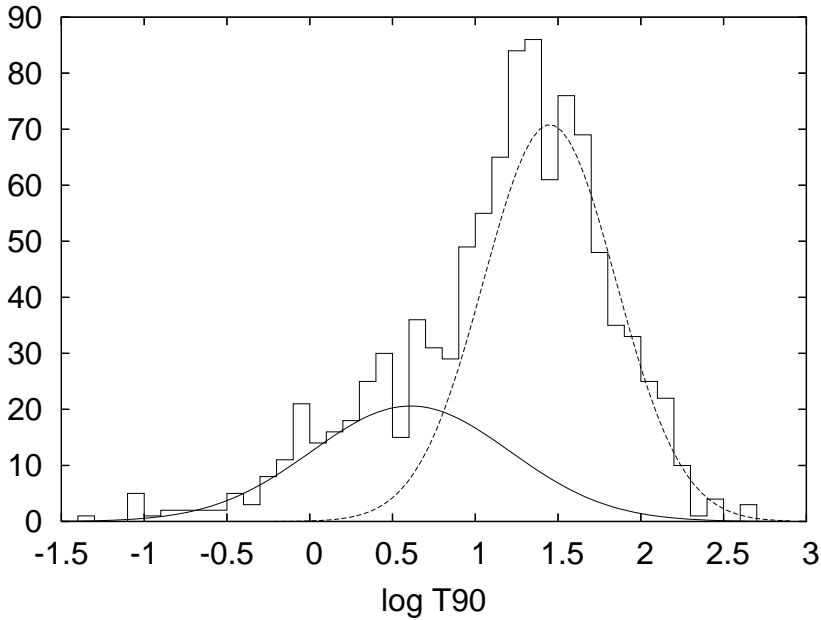}
\caption{The duration distribution of the Beppo-Sax GRBs.}
\label{fig03} 
\end{figure}

\section{The Beppo-Sax sample}

In the Beppo-Sax catalog  \cite{fr09} there are
1082 GRBs, of which 1003 have duration information
 (see Figure 2. for the distribution).
For the analysis of this distribution we used
the Maximum Likelihood (ML) method.
The ML method was not able to identify
all the three subgroups in the Beppo-Sax data at a high
significance level.
However the third group in our analysis was the shortest in
duration. Therefore analyzing the duration distribution
observed by Beppo-Sax GRBM one can find
the long and the intermediate duration population \cite{hor09}. 
The short population can be seen only with low
significance level.

 \section{Acknowledgements}
		This research is supported by Hungarian OTKA grant K077795, 
by OTKA/NKTH A08-77719 and A08-77815 (ZB),
and  by a Bolyai Scholarship (IH).


%
\end{document}